\documentstyle[12pt,epsfig,axodraw,a4]{article} 
\def\bild#1#2{    
        \vspace*{-5mm}
        \begin{center}
        \begin{math}
        \epsfxsize#2cm
        \epsffile{#1}
        \end{math}
        \end{center}  }
\newcommand{\vs}{\vspace{-0.25cm}}
\begin{document} 
\begin{center}
\large{\bf Iterated and irreducible pion-photon exchange in nuclei}

\medskip

N. Kaiser\\

\smallskip

{\small Physik Department T39, Technische Universit\"{a}t M\"{u}nchen,
    D-85747 Garching, Germany}

\end{center}

\bigskip

\begin{abstract}
We calculate the contribution to the nuclear energy density functional which
arises from iterated pion-photon exchange between nucleons. In heavy nuclei, 
this novel charge symmetry breaking interaction leads to an additional binding
of each proton by about 0.2\,MeV. Compared to that the analogous effect from 
irreducible pion-photon exchange is negligibly small. As a possible mechanism 
to resolve the Nolen-Schiffer anomaly we propose the iteration of one-photon 
exchange with an attractive short-range NN-interaction. The corresponding 
energy per proton reads: $\bar E[\rho_p]=(2\alpha/15\pi^2)(\pi^2 -3+6 \ln2)
{\cal A}_{pp}\,k_p^2$ with $\rho_p =k_p^3/ 3\pi^2$ the proton density and
${\cal A}_{pp}\approx 2\,$fm an effective (in-medium) scattering length. Hints
for such a value of ${\cal A}_{pp}$ come from phenomenological Skyrme forces
and from the neutron  matter equation of state.  

 \end{abstract}

\bigskip
PACS: 12.38.Bx, 21.10.Sf, 31.15.Ew.

%To be published in {\it The Physical Review C (2002), Brief Reports}

\bigskip

\bigskip
A classic problem in nuclear structure theory is to understand the binding
energy differences between mirror nuclei (i.e. nuclei with the same mass
number $A=Z+N$ but with the proton number $Z$ and the neutron number $N$
interchanged). If the strong nuclear force is charge symmetric, then these
binding energy differences can be directly related to the well-understood
Coulomb interaction between the protons. However, as shown long ago by Nolen 
and Schiffer \cite{nolen,shlomo} the experimental binding energy differences 
are systematically (by about 7\%) larger than the ones calculated with a charge
symmetric strong interaction and realistic nuclear wave functions  
(reproducing e.g. elastic electron-scattering data). This discrepancy which 
ranges from fairly light up to the heaviest available mirror nuclei is called 
the Nolen-Schiffer anomaly. It is generally agreed that both nuclear 
correlations \cite{bulgac} and a charge-symmetry breaking strong interaction 
are important for its understanding. 

Recently, Brown et al. \cite{brown} have
performed a systematic study of these binding energy differences the mass
region $A\leq 60$ using the Skyrme-Hartree-Fock method. They have found that
the anomaly can be resolved by either dropping the (weakly) attractive Coulomb
exchange term in the nuclear energy density functional (see eqs.(5,7) below) or
by introducing a charge-symmetry breaking delta-force which splits the
proton-proton and neutron-neutron effective s-wave interactions. The strength
of the adjusted charge-symmetry breaking interaction comes out, however, a
factor 3 to 4 larger than expected from some high-precision nucleon-nucleon
potentials (such as AV18 \cite{av18} or CDBonn \cite{cdbonn}). Henley and Krein
\cite{henley} suggested an alternative explanation of the Nolen-Schiffer
anomaly based on the quark substructure of nucleons in the nuclear medium. In
their Nambu-Jona-Lasinio model the down-up-quark mass difference induces, 
driven by the partial restoration of chiral symmetry, a substantial reduction
of the neutron-proton mass difference in the nuclear medium (see Fig.\,2b in
ref.\cite{henley}). Clearly, any decrease of the neutron-proton mass difference
will help to resolve the Nolen-Schiffer anomaly, since the calculated binding
energy differences of mirror nuclei are based on the free neutron-proton mass
difference of $1.293\,$MeV. For a similar approach using the quark-meson 
coupling model, see ref.\cite{thomas}. In this context it should also be noted 
that most relativistic (scalar-vector) mean-field models for nuclear structure
leave out the (attractive) Coulomb exchange term by default and thus circumvent
the problem.  

In this work we will use chiral perturbation theory to calculate the leading
order long-range charge-symmetry breaking effects generated by the combined 
pion- and photon-exchange between nucleons. We present analytical results for 
the corresponding contributions to the nuclear energy density functional. This
energy density functional actually defines a general starting point for 
(non-relativistic) nuclear structure calculations within the self-consistent
mean-field approximation \cite{reinhard}. It turns out that the three-loop 
in-medium diagrams of iterated pion-photon exchange and of irreducible 
pion-photon exchange contribute at different orders in the small momentum
expansion. This allows us even to draw some conclusions about convergence 
properties.  

\medskip

\bild{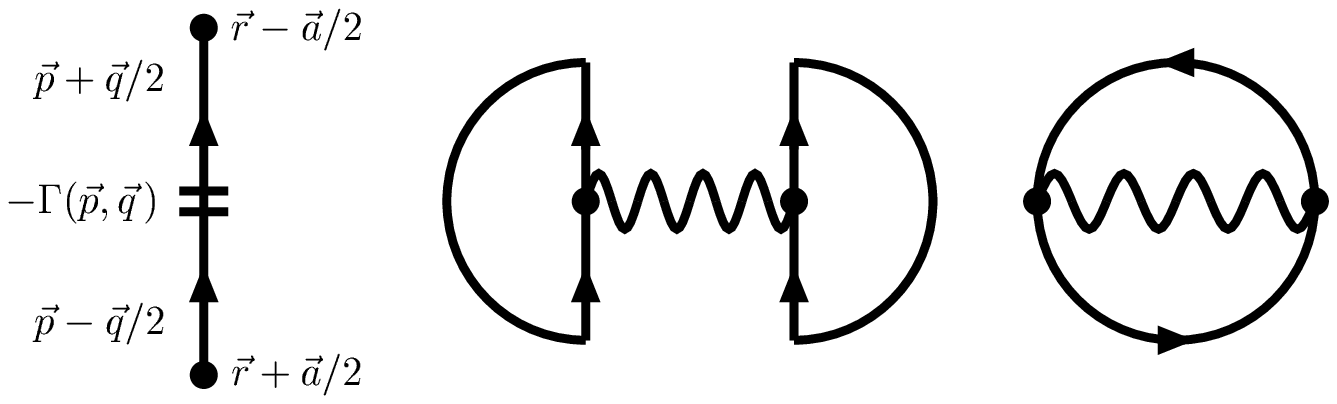}{12}
\vspace{-0.5cm}
{\it Fig.\,1: Left: The double line symbolizes the medium insertion $\Gamma(
\vec p,\vec q\,)$ defined by eq.(4). Next shown are the two-loop one-photon 
exchange Hartree and Fock diagrams. Their combinatoric factor is 1/2.} 

\medskip

The starting point for the construction of an explicit nuclear energy density
functional is the density-matrix as given by a sum over the occupied energy
eigenfunctions $\Psi_\nu$. We restrict here the formulation to the 
proton-states since only these participate in the Coulomb interaction. 
According to Negele and Vautherin \cite{negele} the bilocal density-matrix can
be expanded in relative and center-of-mass coordinates, $\vec a$ and $\vec r$,
as follows:     
\begin{eqnarray} \sum_{\nu\in \rm occ}\Psi_\nu( \vec r -\vec a/2)
\Psi_\nu^\dagger(\vec r +\vec a/2) &=& {3 \rho_p\over a k_p}\, j_1(a k_p)-
{35 \over 2a k_p^3} \,j_3(a k_p) \bigg[ \tau_p - {3\over 5} \rho_p k_p^2 -
{1\over 4} \vec \nabla^2 \rho_p \bigg]\,,  \end{eqnarray}
where $j_{1,3}(a k_p)$ are ordinary spherical Bessel functions. The other 
quantities on the right hand side of eq.(1) are the local proton density:
\begin{equation} \rho_p(\vec r\,) =  {k_p^3(\vec r\,)\over 3\pi^2} = \sum_{
\nu\in \rm occ}\Psi^\dagger_\nu( \vec r\,)\Psi_\nu( \vec r\,)\,,\end{equation}
written in terms of a local proton Fermi-momentum $k_p(\vec r\,)$, and the
local proton kinetic energy density:
\begin{equation} \tau_p(\vec r\,) = \sum_{\nu\in\rm occ}\vec \nabla\Psi^\dagger
_\nu( \vec r\,)\cdot \vec \nabla\Psi_\nu( \vec r\,) \,.\end{equation}
The Fourier transform of the expanded density-matrix in eq.(1) with respect to
both coordinates $\vec a$ and $\vec r$ defines a "medium insertion" for the
inhomogeneous many-fermion system \cite{efun}:
\begin{equation} \Gamma(\vec p,\vec q\,) = \int d^3 r \, e^{-i \vec q \cdot
\vec r}\,\theta(k_p-|\vec p\,|)\, \bigg\{1 +{35 \pi^2 \over 4k_p^7}(5\vec
p\,^2 -3k_p^2) \bigg[ \tau_p - {3\over 5} \rho_p k_p^2 - {1\over 4} \vec 
\nabla^2 \rho_p \bigg] \bigg\}\,.  \end{equation}
The double line in Fig.\,1 symbolizes this medium insertion together with the
assignment of the out- and in-going momenta $\vec p \pm \vec q/2$. The momentum
transfer $\vec q$ is provided by the Fourier components of the inhomogeneous
(proton) distributions $\rho_p(\vec r\,)$ and $\tau_p(\vec r\,)$. Going up to
second order in spatial gradients (i.e. deviations from homogeneity) the
protonic energy density functional reads:
\begin{equation} {\cal E}[\rho_p,\tau_p] =\rho_p\bar E[\rho_p]+\bigg[\tau_p-
{3\over 5} \rho_p k_p^2- {1\over 4} \vec \nabla^2 \rho_p\bigg] F_\tau(\rho_p) 
\,. \end{equation}
Here, the functional $\bar E[\rho_p]$ denotes the energy per proton. The other 
strength function $F_\tau(\rho_p)$ is related to the effective proton mass by 
the relation: $\widetilde M^*(\rho_p)= M[1+2M F_\tau(\rho_p)]^{-1}$, where 
$M=938.272\,$MeV denotes the free proton mass. We apply the density-matrix 
formalism first to the one-photon exchange diagrams in Fig.\,1. From the 
Hartree diagram one regains the well-known classical result: 
\begin{equation} \bar E[\rho_p]= {\alpha\over 2} \int\! d^3r'{\rho_p(\vec r\,')
\over |\vec r - \vec r\,'|} = {2\pi \alpha \over r } \int_0^\infty\! dr'r' 
\rho_p(r')  {\rm min}(r,r')\,, \end{equation} 
with $\alpha = 1/137.036$ the fine structure constant. The simplification of
expression for the Coulomb potential in the second part of eq.(6) holds for a 
spherically symmetric proton density $\rho_p(r')$. The result of the one-photon
exchange Fock diagram is also well-known \cite{titin}:
\begin{equation} \bar E[\rho_p]= -{3\alpha\over 4\pi} \, k_p \,, \qquad \qquad
F_\tau(\rho_p)= {35\alpha\over 36\pi k_p} \,. \end{equation}
The derivative of the corresponding energy density with respect to the proton 
density:\\ $\partial (\rho_p\bar E[\rho_p])/\partial \rho_p = -\alpha k_p/\pi$,
is sometimes also referred to as the Coulomb exchange term \cite{brown}. 

\bigskip
\medskip

\bild{pigamma.epsi}{12}

{\it Fig.\,2: Iterated pion-photon exchange Hartree and Fock diagrams.
Their combinatoric factor is 1/2.}

\bigskip

Next, we consider pion-loop corrections to the one-photon exchange diagrams in
Fig.\,1. There are many possible diagrams, some introduce vacuum polarization, 
others generate (contributions to) the proton electric form factor and another
class builds up the $\pi\gamma$-exchange nucleon-nucleon potential (see eq.(16)
below). The two diagrams of iterated pion-photon exchange shown in Fig.\,2 are 
however special and distinguished by the feature of carrying an energy 
denominator equal to the difference of nucleon kinetic energies. These two 
diagrams get enhanced by the large scale factor $M$ and therefore they are 
expected to be dominant. The left Hartree diagram in Fig.\,2 is easily seen to 
vanish as a consequence of a zero spin-trace, tr$(\vec \sigma\cdot\vec l\,)=0$.
The evaluation of the iterated $\pi\gamma$-exchange Fock diagram with two 
medium insertions (on non-neighboring nucleon propagators) leads to the loop 
integral:  
\begin{eqnarray}&& -\!\!\!\!\!\!\int {d^3l\over 2\pi^2} 
{(\vec l +\vec Q)^2\over [m_\gamma^2 +\vec l\,^2] \,[m_\pi^2+(\vec l +\vec Q)^2
] \,\vec l \cdot( \vec l+\vec Q)}= {1\over Q (m_\pi^2+
m_\gamma^2+Q^2)} \nonumber \\ && \times \Bigg[ (m_\gamma^2+Q^2)\arctan{Q\over 
m_\gamma} -m_\pi^2\arctan{Q\over m_\pi}+2m_\pi^2\arctan{Q\over m_\pi+m_\gamma}
\Bigg] \,, \end{eqnarray} 
which has been regularized by introducing an infinitesimal photon mass 
$m_\gamma$. Taking the limit $m_\gamma \to 0$ of the analytical expression in
eq.(8) one gets the following infrared-finite result for the loop integral:
\begin{equation} {1 \over m_\pi^2+Q^2}\Bigg[ {\pi \over 2} |Q|+{m_\pi^2 \over 
Q} \arctan{Q\over m_\pi} \Bigg] \,. \end{equation}
We warn that by setting $m_\gamma =0$ in the loop integrand and performing 
some seemingly harmless manipulations one can arrive at an incorrect result 
which misses the $\pi|Q|/2$ term in eq.(9). After further reduction of
the integral over the product of two Fermi spheres of radius $k_p$ one obtains
the following contribution to the energy per proton:  
\begin{equation} \bar E[\rho_p]= -{\alpha g_{\pi N}^2 m_\pi^2 \over 4\pi^2 M
u^3}\int_0^u\! dx\,x(u-x)^2(2u+x){2\pi x^2 +\arctan 2x \over 1+4x^2} \,, 
\end{equation}
with the abbreviation $u = k_p/m_\pi$. We choose the value $g_{\pi N}=13.2$ for
the pion-nucleon coupling constant and $m_\pi = 135\,$MeV stands for the 
neutral pion mass. In order to recover the enhancement factor $M$ in eq.(10) 
one has to remember that the pseudovector $\pi NN$-vertex carries the
prefactor $g_{\pi N}/2M$. The contribution to the strength function
$F_\tau(\rho_p)$ reads:  
\begin{equation} F_\tau(\rho_p)= {35\alpha g_{\pi N}^2 \over 12\pi^2 M u^7} 
\int_0^u\! dx \,x^2(u-x)^2(3u^2-4ux-2x^2) {2\pi x^2 +\arctan 2x \over 1+4x^2}  
\,, \end{equation}
where we have made use of the master integral eq.(A1) in ref.\cite{efun}. 
An in-medium diagram with three medium insertions represents Pauli-blocking
effects. We obtain from the iterated  $\pi\gamma$-exchange Fock diagram with
three medium insertions the following contributions to the energy per proton: 
\begin{equation} \bar E[\rho_p]= {3\alpha g_{\pi N}^2 m_\pi^2 \over 16\pi^3 M
u^3} \int_0^u \!dx\,x^2\!\int_{-1}^1\!dy \!\int_{-1}^1 \!dz\, {yz \,\theta(y^2+
z^2-1) \over |yz|\sqrt{y^2+z^2-1}}\Big[s^2- \ln(1+s^2)\Big]\ln t
\,,\end{equation} 
and to the strength function $F_\tau(\rho_p)$:
\begin{eqnarray} F_\tau(\rho_p)&=& {175\alpha g_{\pi N}^2\over 64\pi^3 M u^7} 
\int_0^u \!dx\,x^2\!\int_{-1}^1\!dy \!\int_{-1}^1 \!dz\,{yz\,\theta(y^2+z^2-1) 
\over |yz|\sqrt{y^2+z^2-1}}\Bigg\{\Big[\ln(1+s^2)-s^2 \Big] \nonumber \\ &&
\times \bigg[tx z +\bigg({9\over 5} u^2 +1-3x^2 \bigg) \ln t \bigg]+ \bigg[4xy
\bigg(s-{s^3\over 3}-\arctan s\bigg)+{s^4\over 2}\bigg]\ln t \Bigg\}\,.
\end{eqnarray} 
Here, $s=xy +\sqrt{u^2-x^2+x^2y^2}$ and $t=xz +\sqrt{u^2-x^2+x^2z^2}$ are
auxiliary functions which emerge in the reduction \cite{efun} of a 
nine-dimensional principal value integral over the product of three Fermi
spheres of radius $k_p$. The infrared finiteness of the results eqs.(12,13) 
deserves some closer inspection. In the actual calculation the logarithmic
term $\ln t$ is accompanied by an additive constant $\ln(m_\pi/m_\gamma)$. 
However, when integrated over $z$ the infrared-singular constant $\ln(m_\pi/
m_\gamma)$ drops out since the integrand is odd under the substitution $z\to 
-z$.  It is also interesting to consider the chiral limit of vanishing pion
mass $m_\pi=0$. In that case all integrals occurring in eqs.(10-13) can be
solved in closed form. One finds a simple $\rho_p^{2/3}$-behavior of the energy
per proton: 
\begin{equation} \bar E[\rho_p]\Big|_{m_\pi=0} ={\alpha g_{\pi N}^2 k_p^2 \over
120\pi^3 M} \Big(3-\pi^2 -6 \ln2\Big) \simeq -1.50\,{\rm MeVfm}^2 \cdot
\rho_p^{2/3} \,, \end{equation}
and a density independent strength function $F_\tau(\rho_p)$: 
\begin{equation}  F_\tau(\rho_p)\Big|_{m_\pi=0} ={\alpha g_{\pi N}^2 
\over \pi^3 M} \bigg({13\over 64}-{2\over 3}\ln2\bigg) \simeq -0.441\,{\rm MeV
fm}^2 \,.\end{equation}
While these density dependences could also be guessed through mass dimension
counting, the numerical coefficients are highly non-trivial.

\bigskip

\bild{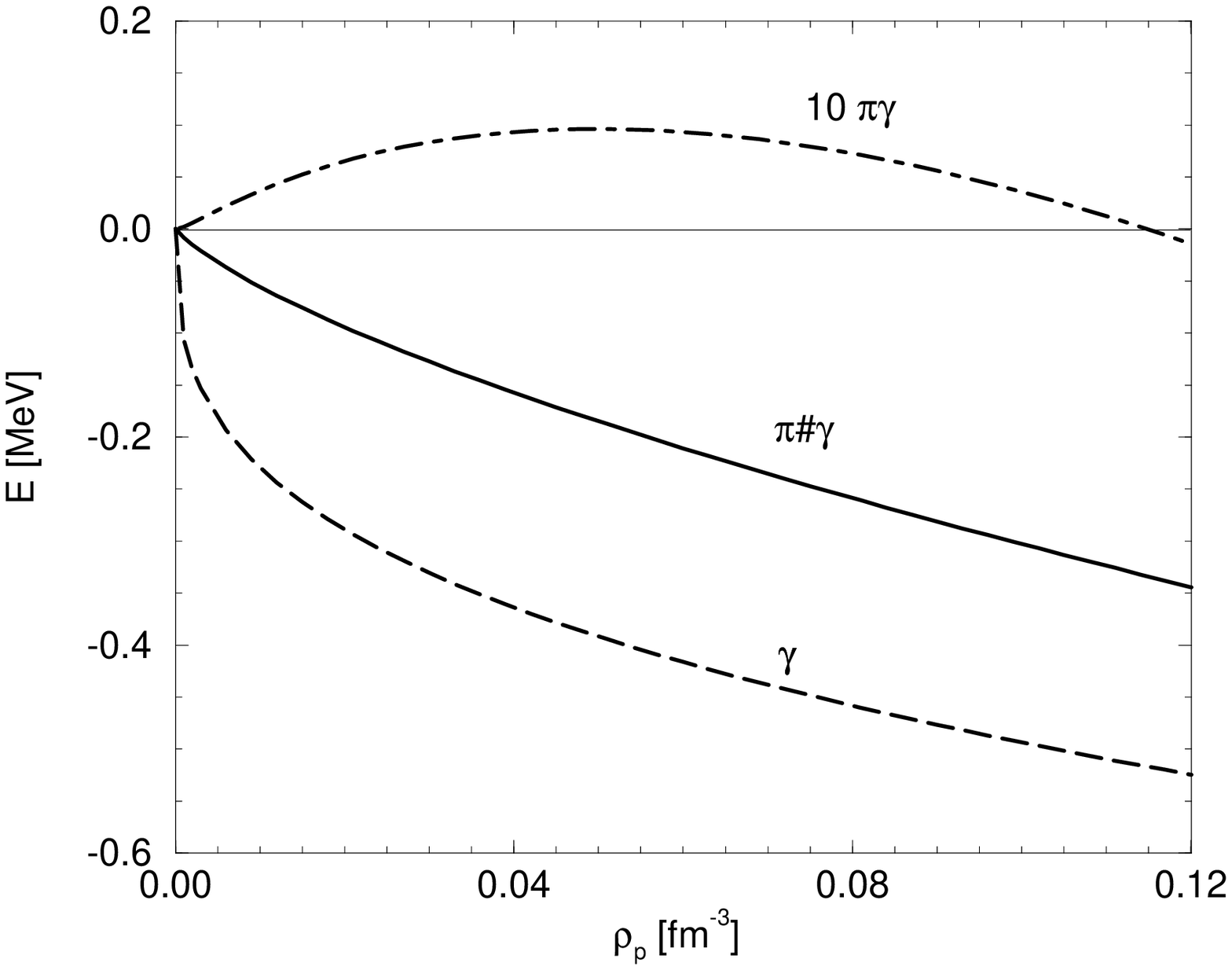}{14}
\vskip -1.cm
{\it Fig.\,3: The energy per proton $\bar E[\rho_p]$ versus the proton density
$\rho_p = k_p^3/3\pi^2$. The dashed and full line show the result of the 
$1\gamma$-exchange and the iterated $\pi\gamma$-exchange Fock diagram. The 
dashed-dotted line shows the result of irreducible $\pi\gamma$-exchange, 
magnified by a factor 10.}

\bigskip

In Fig.\,3 we show the energy per proton $\bar E[\rho_p]$ as a function of the
proton density $\rho_p = k_p^3/3\pi^2$. The dashed line shows the result $\bar
E[\rho_p] = -3\alpha k_p/4\pi$ of the one-photon exchange Fock diagram. The
full line gives the result of the iterated pion-photon exchange Fock diagram
(evaluated with a finite pion mass of $m_\pi= 135\,$MeV). One notices that both
energy densities lead in nuclei to an attraction between the protons. 
Furthermore, as it should be for a higher order correction, iterated
$\pi\gamma$-exchange comes out a factor of about two smaller than  
$1\gamma$-exchange. The density dependence of the full line in Fig.\,3 is well
approximated by the fit function: $\bar E[\rho_p]^{\pi\gamma}\simeq
-1.4\,$MeVfm$^2\cdot\rho_p^{2/3}$. Interestingly, it does not differ much from
the result eq.(14) valid in the strict chiral limit $m_\pi=0$.

\bigskip

\bild{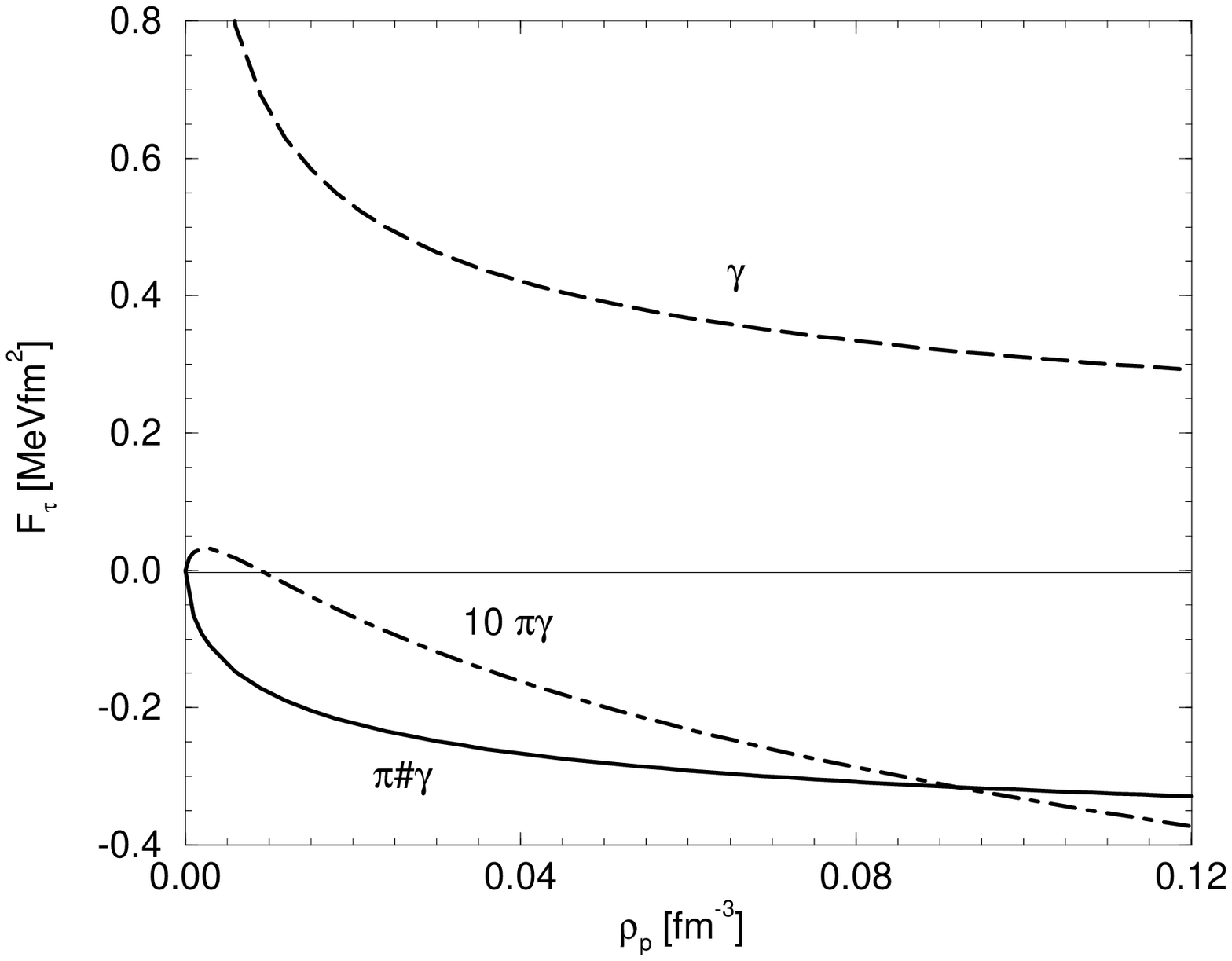}{14}
\vskip -1.cm
{\it Fig.\,4: The strength function $F_\tau(\rho_p)$ versus the proton density
$\rho_p = k_p^3/3\pi^2$. The dashed and full line show the result of the 
$1\gamma$-exchange and the iterated $\pi\gamma$-exchange Fock diagram. The 
dashed-dotted line shows the result of irreducible $\pi\gamma$-exchange, 
magnified by a factor 10.}

\bigskip

In Fig.\,4 we show the strength function $F_\tau(\rho_p)$ versus the proton
density $\rho_p = k_p^3/3\pi^2$. The dashed line shows the result $F_\tau(
\rho_p) = 35\alpha/36\pi k_p$ of the one-photon exchange Fock diagram and the 
full line gives the result of the iterated pion-photon exchange Fock diagram. 
One observes a strong cancelation between both contributions, in particular at 
higher proton densities $\rho_p > 0.06\,$fm$^{-3}$. Therefore one may neglect 
with good reason the part of the energy density functional proportional to 
$F_\tau(\rho_p)$ in nuclear structure calculations.  

For a numerical estimate of the Coulomb energies of nuclei we choose radially 
symmetric proton densities of Saxon-Woods type: $\rho_p(r) \sim [1+\exp((r-R)/
a)]^{-1}$, with a diffuseness parameter of $a=0.54\,$fm and a mean nuclear 
radius of $R=1.07\,$fm$\cdot A^{1/3}$ (where $A$ is the mass number). The 
proton density $\rho_p(r)$ of a given nucleus is normalized to its total proton
number $Z= 4\pi\int_0^\infty dr r^2 \rho_p(r)$ and the Coulomb energy is then
calculated via: $E_{\rm Coul}=4\pi\int_0^\infty drr^2\rho_p(r)\bar 
E[\rho_p(r)]$. In Tab.\,1 we list numerical values of the Coulomb energies per
proton $E_{\rm Coul}/Z$ for a few selected nuclei: Ca, Fe, Sn and Pb. The 
direct Coulomb term (labeled $1\gamma$-Hartree) gives of course the dominant 
repulsive contribution which increases with the proton number $Z$. The Coulomb 
exchange term (labeled $1\gamma$-Fock) is weakly attractive and almost 
independent of the proton number $Z$. The same features hold for the iterated 
$\pi\gamma$-exchange Fock diagram which gives numerical values of $E_{\rm Coul}
/Z$ about half as large as those of the Coulomb exchange term. One can infer 
from Tab.\,1 that iterated $\pi\gamma$-exchange leads to an additional binding 
of each proton in a heavy nucleus by about 0.2\,MeV. This unique long-range 
charge-symmetry breaking interaction does therefore not help to resolve the 
Nolen-Schiffer anomaly. On the contrary, it even further enhances the 
discrepancies.  

\bigskip

\begin{table}[hbt]
\begin{center}
\begin{tabular}{|c|cccc|}
    \hline
    $E_{\rm Coul}/Z$ [MeV] & Ca & Fe & Sn & Pb \\ \hline  $1\gamma$-Hartree 
& 3.94 & 4.72 & 7.44 & 10.43 \\  $1\gamma$-Fock & --0.37 &--0.37 &--0.38
&--0.39  \\ $\pi\gamma$-Fock & --0.18 &--0.18 &--0.19 &--0.20 \\
  \hline
  \end{tabular}
\end{center}

{\it Tab.1: Numerical values of Coulomb energies per proton for various
nuclei. The units of $E_{\rm Coul}/Z$ are MeV.}
\end{table}
\bigskip

\bigskip

\bild{pgirr1.epsi}{12}
\bild{pgirr2.epsi}{12}
{\it Fig.\,5: Fock diagrams of irreducible pion-photon exchange. Their 
combinatoric factors are 1, 1/2, 1/2, 1, 1 and 1/2, in the order shown.}

\bigskip

Next, we consider irreducible pion-photon exchange which is of subleading order
in the small momentum expansion (no large scale enhancement factor $M$ is 
present). The relevant set of Fock diagrams contributing to the energy density
functional is shown in Fig.\,5. The corresponding Hartree diagrams (with two
closed nucleon lines) vanish again due to a zero spin-trace. By opening two
nucleon lines of the diagrams in Fig.\,5 one encounters the NN-scattering
T-matrix related to irreducible $\pi\gamma$-exchange between nucleons. This 
T-matrix has recently been calculated in chiral perturbation theory by van
Kolck et al. \cite{kolck}. We reproduce independently their analytical result: 
\begin{equation} T_{\pi \gamma}(\vec Q)  = {\alpha g_{\pi N}^2 \over 8\pi M^2}
(\vec\tau_1 \cdot\vec\tau_2-\tau_1^3 \tau_2^3) \, \vec \sigma_1 \cdot\vec Q \, 
\vec \sigma_2 \cdot \vec Q \Bigg\{ {1\over Q^2} - {(m_\pi^2- Q^2)^2 \over Q^4 
(m_\pi^2+ Q^2)} \ln\bigg(1+ {Q^2\over m_\pi^2}\bigg) \Bigg\}\,. \end{equation} 
Here, $\vec \sigma_{1,2}$ and $\vec \tau_{1,2}$ denote the spin- and 
isospin-operators of the two nucleons and $\vec Q$ is the momentum transfer
between both nucleons. Obviously, the charge-symmetry breaking T-matrix in 
eq.(16) can only contribute to elastic $pn\to np$ scattering. The corresponding
energy density depends therefore on the product of the proton and the neutron
density. For the sake of simplicity we assume these densities to be equal and 
obtain then the following contributions from irreducible $\pi\gamma$-exchange
to the energy per proton:  
\begin{equation} \bar E[\rho_p]= {\alpha g_{\pi N}^2 m_\pi^3 \over 8\pi^3 M^2}
\Bigg\{ {u^3 \over 3}-\int_0^u\! dx\,{(u-x)^2(2u+x)\over u^3(1+4x^2)}(1-4x^2)^2
\ln(1+4x^2) \Bigg\} \,, \end{equation}
and to the strength function $F_\tau(\rho_p)$:
\begin{equation} F_\tau(\rho_p)= {35\alpha g_{\pi N}^2 m_\pi\over 24\pi^3 M^2 
u^7} \int_0^u\! dx \,x(u-x)^2(3u^2-4ux-2x^2) {(1-4x^2)^2\over 1+4x^2}
\ln(1+4x^2) \,. \end{equation}
The density dependence of these results (magnified by a factor of 10) is shown 
by the dashed-dotted lines in Figs.\,3,4. One observes that the contributions 
from irreducible $\pi\gamma$-exchange are negligibly small. The Coulomb
energies per proton $E_{\rm Coul}/Z$ are affected at the level of a few
keV. Such tiny repulsive corrections can of course not help to resolve the
Nolen-Schiffer anomaly. For the sake of completeness we consider also the
behavior of the expressions in eqs.(17,18) in the chiral limit, $m_\pi \to 0$,
and find: 
\begin{equation} \bar E[\rho_p]\Big|_{m_\pi\to 0} ={\alpha g_{\pi N}^2 k_p^3 
\over 12\pi^3 M^2} \bigg({5\over 4}-\ln{2k_p\over m_\pi}\bigg) \,, \qquad  
F_\tau(\rho_p)\Big|_{m_\pi=0} =-{7\alpha g_{\pi N}^2 k_p \over 96\pi^3 M^2}\,.
\end{equation}
In order to resolve the Nolen-Schiffer anomaly a dynamical mechanism is needed
which generates additional repulsion for the protons but leaves the neutrons
unaffected. Inspired by eq.(14), we propose as such a possible mechanism the 
iteration of a short-range attractive NN-interaction with the (repulsive) 
one-photon exchange. For its explicit evaluation consider the three-loop 
diagrams in Fig.\,2 with the dashed line symbolizing now a heavy scalar meson. 
The ratio $g_s/m_s$ of the scalar meson's coupling constant and mass merely
serves to introduce an effective scattering length ${\cal A}_{pp}=(g_s/m_s)^2
M/4\pi$. After performing all occurring integrals we obtain from the iteration
of a short-range $pp$-interaction with one-photon exchange the following
contributions to the energy per proton: 
\begin{equation} \bar E[\rho_p]={2\alpha \over 15\pi^2}(\pi^2 -3+6 \ln2) 
{\cal A}_{pp}\, k_p^2\simeq 2.053\,{\rm MeVfm}\cdot {\cal A}_{pp}\,\rho_p^{2/3}
\,, \end{equation}
and to the strength function $F_\tau(\rho_p)$:
\begin{equation}  F_\tau(\rho_p)={\alpha \over \pi^2} \bigg({32\over 3}\ln2-
{13\over 4}\bigg) {\cal A}_{pp} \,.\end{equation}
The sign-convention is chosen here such that a positive scattering length 
${\cal A}_{pp}>0$ corresponds to attraction. 

In order to compensate both the 
$1\gamma$-exchange Fock contribution and the $\pi\gamma$-exchange Fock 
contribution an effective $pp$-scattering length of ${\cal A}_{pp}\approx 
2$\,fm is required. This numerical estimate stems from the numbers in the
second and third row of Tab.\,1 in combination with eq.(20) and the fit
function $\bar E[\rho_p]^{\pi\gamma} \simeq -1.4\,$MeVfm$^2\cdot\rho_p^{2/3}$. 
Note that this estimated value of the effective $pp$-scattering length ${\cal
A}_{pp} \approx 2$\,fm is an order of magnitude smaller than the free $^1S_0$ 
NN-scattering length $a(^1S_0) \simeq 19\,$fm \cite{trotter} and therefore it 
subsumes lots of many-body dynamics being active at finite density. Most 
interestingly, phenomenological Skyrme forces give a strong indication for such
a value of the effective $pp$-scattering length. As a matter of fact, the 
Skyrme force parameters $t_0$ and $x_0$ \cite{brown,reinhard} encode an 
effective $pp$-scattering length via the relation ${\cal A}_{pp}=M t_0(x_0-1)
/4\pi$. Inserting the values in Tab.\,1 of ref.\cite{skx} one finds ${\cal 
A}_{pp}= 1.96\,$fm, $2.40\,$fm and $1.83\,$fm for the forces SKXce, SKXm and
SKX, respectively. Averaging over the variants MSk1-6 of ref.\cite{pearson} one
gets for comparison ${\cal A}_{pp} =1.66\,$fm. A further hint on ${\cal
A}_{pp}$ comes from the equation of state of pure neutron matter. The result of
the sophisticated many-body calculation of the Urbana group \cite{urbana} is
well approximated for neutron densities $\rho_n  \leq 0.4\,$fm$^{-3}$ by a
fourth order polynomial in the neutron Fermi momentum $k_n$, namely: $\bar
E_n(k_n) =3k_n^2/10M-\alpha_n k_n^3/M^2+\beta_n k_n^4/M^3$, with adjusted
coefficients $\alpha_n =1.247$ and $\beta_n=2.168$. The coefficient $\alpha_n$
of the term linear in the neutron density $\rho_n = k_n^3/3\pi^2$ can be 
translated into an effective $nn$-scattering length of ${\cal A}_{nn}= 3\pi 
\alpha_n/M \simeq 2.47\,$fm and isospin symmetry implies ${\cal A}_{pp}= {\cal
A}_{nn}\simeq 2.47\,$fm. Admittedly, these arguments for ${\cal A}_{pp} \approx
2\,$fm leave some loose ends but there seem to be several interesting cross 
relations. 

In summary, we have calculated in this work the contributions of iterated and
irreducible pion-photon exchange to the nuclear energy density functional. 
These novel terms represent unique long-range charge-symmetry breaking 
interactions. In heavy nuclei, iterated $\pi\gamma$-exchange leads to an
additional binding of each proton by about $0.2\,$MeV. As a consequence of that
the discrepancies between calculated and experimental binding energy
differences of mirror nuclei will be further enhanced. The subleading effects 
from irreducible $\pi\gamma$-exchange turn out to be negligibly small. As a
possible mechanism to resolve the Nolen-Schiffer anomaly we propose the
iteration of a short-range attractive NN-interaction with one-photon
exchange. The corresponding energy per proton has the simple analytical form:
$\bar E[\rho_p]=(2\alpha/15\pi^2)(\pi^2-3+6\ln2) {\cal A}_{pp}\,k_p^2$. There
are various hints from phenomenological Skyrme forces and from the neutron 
matter equation of state for an effective $pp$-scattering length with a value 
around ${\cal A}_{pp} \approx 2$\,fm. In a next (more quantitative) step one 
should perform nuclear structure calculations (e.g. within the
Skyrme-Hartree-Fock approach) which include all the charge-symmetry breaking
terms derived in this work. The value of the effective scattering length ${\cal
A}_{pp}$ should be determined in a best fit to the binding energy differences
of mirror nuclei, and one should check the relation  ${\cal A}_{pp}=M t_0(x_0
-1)/4\pi$ to the Skyrme force parameters $t_0$ and $x_0$. Of course, a more 
microscopic interpretation/derivation of it would also be desirable.  

\end{document}